\documentclass[
a4paper,twocolumn,11pt,accepted=2026-10-02]{quantumarticle}
\pdfoutput=1

\usepackage{babel}
\usepackage[utf8]{inputenc}
\usepackage[T1]{fontenc}
\usepackage{graphicx}   
\usepackage{dcolumn}    
\usepackage{bm}         
\usepackage{braket}     
\usepackage{comment}    
\usepackage[dvipsnames]{xcolor}
\usepackage{hyperref}
\usepackage{ragged2e}
\usepackage{enumitem}
\usepackage{amsmath,amssymb}

\usepackage{tcolorbox}
\tcbuselibrary{listings, breakable, skins, theorems}

\newtcolorbox[auto counter]{algorithm}[2][]{%
  enhanced,
  float=!ht,
  colback=white,
  colframe=black,
  fonttitle=\bfseries,
  title=Algorithm~\thetcbcounter: #2,
  label={#1},
  boxrule=0.5pt,
  arc=2pt,
  left=6pt,
  right=6pt,
  top=6pt,
  bottom=6pt
}

\newcounter{ga}

\usepackage{xcolor}

\graphicspath{ {./images/} }

\begin{document}
\title{Quantum Walks on Arbitrary Spatial Networks with Rydberg Atoms}

\author{Gabriel Almeida}
\email{gabriel.m.almeida@tecnico.ulisboa.pt}
\affiliation{Instituto Superior Técnico, Universidade de Lisboa, Portugal}
\affiliation{PQI -- Portuguese Quantum Institute, Portugal}

\author{Raul Santos}
\affiliation{Eindhoven University of Technology, Netherlands}

\author{Lara Janiurek}
\affiliation{PQI -- Portuguese Quantum Institute, Portugal}
\affiliation{University of Strathclyde, Scotland, United Kingdom}

\author{Yasser Omar}
\email{contact.yasser@pqi.pt}
\affiliation{Instituto Superior Técnico, Universidade de Lisboa, Portugal}
\affiliation{PQI -- Portuguese Quantum Institute, Portugal}
\affiliation{Physics of Information and Quantum Technologies Group, Centro de Física e Engenharia de Materiais Avançados (CeFEMA), Portugal}

\begin{abstract}
Rydberg atoms provide a highly promising platform for quantum computation, leveraging their strong tunable interactions to encode and manipulate information in the electronic states of individual atoms. Key advantages of Rydberg atoms include scalability, reconfigurable connectivity, and native multi-qubit gates, making them particularly well-suited for addressing complex network problems. These problems can often be framed as graph-based tasks, which can be efficiently addressed using quantum walks. In this work, we propose a general implementation of staggered quantum walks with Rydberg atoms, with a particular focus on spatial networks. We also present an efficient algorithm for constructing the tessellations required for the staggered quantum walk. Finally, we demonstrate that our proposal achieves performance consistent with a quadratic speedup in spatial search algorithms.
\end{abstract}
\maketitle

\section{Introduction}
From our social media connections to the layout of public transportation systems and the folded structures of proteins within our cells, complex networks permeate many aspects of the natural and human world \cite{posfai2016network}. These networks, often modeled as graphs, provide a unifying framework for representing diverse systems. Consequently, solving problems on networks --- such as community detection \cite{fortunato2010community}, optimal routing \cite{danila2006optimal,yan2006efficient}, or information propagation \cite{moreno2004dynamics} --- has long captivated scientists across disciplines for both its theoretical richness and its wide-ranging practical applications. However, classical algorithms often face efficiency limitations, particularly as graph sizes grow or the network structures become more complex.

Quantum computing has emerged as a promising approach to solve complex problems faster than classical computers. Several quantum algorithms have been proven to offer a computational advantage, and in the past decades, significant experimental efforts have been made to develop quantum hardware capable of implementing these algorithms. One promising platform is Rydberg atom quantum computing, where quantum information is encoded in the electronic states of atoms, and quantum gates are realized through atom-light interactions and strong Rydberg interactions. This platform, besides promising scalability, offers key advantages, namely reconfigurable connectivity and native multiqubit gates \cite{morgado2021quantum,mcinroy2024benchmarking, maskara2025programmable}. These properties make Rydberg atom systems particularly well-suited for solving network problems. Recent demonstrations with Rydberg atoms include the maximum independent set problem \cite{ebadi2022quantum, kim2022rydberg}, satisfiability problems \cite{jeong2023quantum}, and spatial search problems \cite{young2022tweezer}. 

Other quantum computing platforms have also been explored for solving graph-based problems. In superconducting qubits, QAOA (Quantum Approximate Optimization Algorithm) has been applied to the Sherrington–Kirkpatrick model and the MaxCut problem, but experiments have shown that performance tends to decrease with problem size when the problem graph is not hardware-native \cite{harrigan2021quantum}. In contrast, trapped ion systems have demonstrated more favorable scaling, with QAOA performance improving as the number of rounds increases, even for arbitrary graphs \cite{zhu2022multi}. In photonic processors, Gaussian bosonic samples can be used to solve non-planar graph problems like the Max-Haf problem and the dense $k$-subgraph problem \cite{deng2023solving}. Besides this progress, a general scalable approach to tackle network problems remains a difficult task.

\begin{figure}[b]
    \centering
    \includegraphics[width=0.9\linewidth]{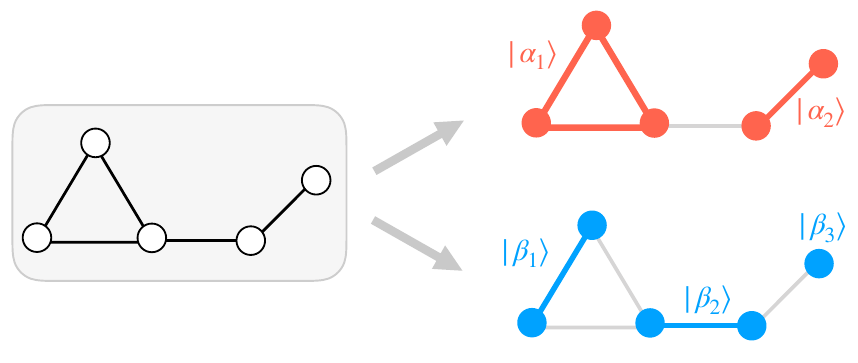}
    \caption{\justifying To perform a staggered walk, a tessellation cover is needed. Each tessellation ($\alpha, \beta,\dots$) is a partition of the vertices into cliques and every edge should belong to at least one tessellation. Each clique is associated with a superposition state of its vertices, from which the walk operator is constructed.}
    \label{fig:tessellation-cover}
\end{figure}

\begin{figure*}[t]
    \centering
    \includegraphics[width=0.98\linewidth]{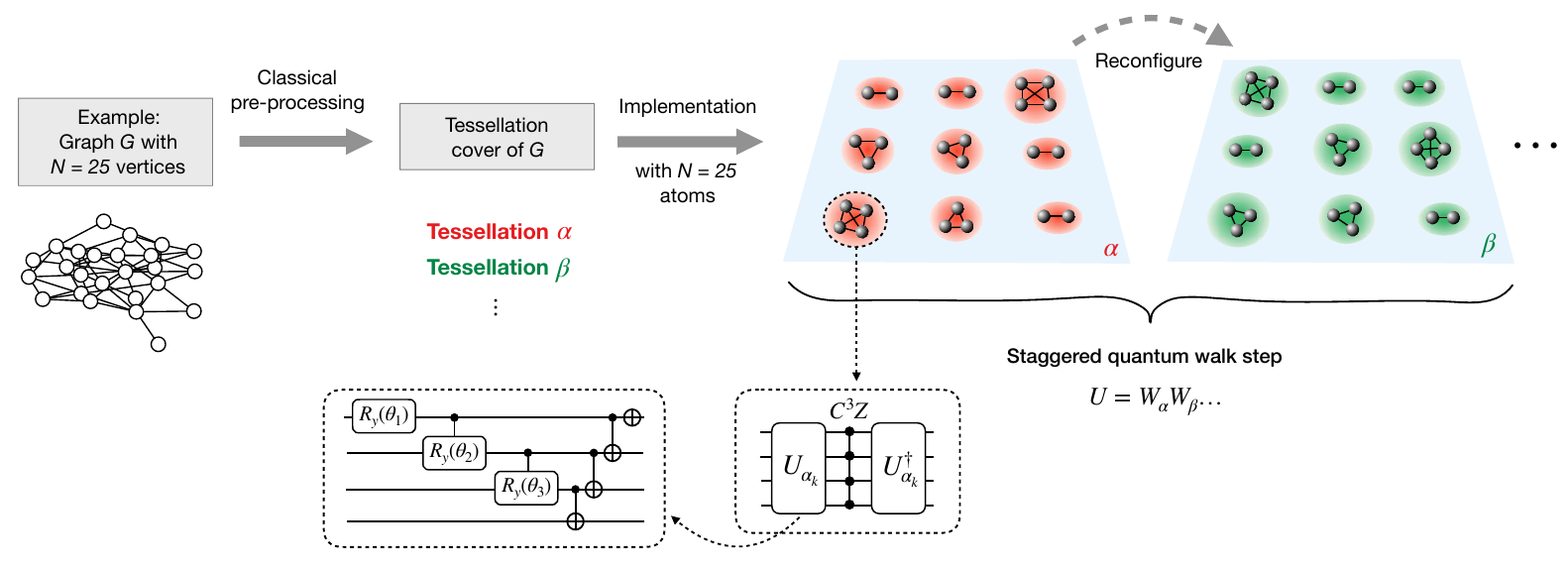}
    \caption{Proposed implementation of a staggered quantum walk. Each vertex of the spatial network is encoded in a single atom and for each tessellation ($\alpha, \beta, \dots$), the walk operator is diagonalized on each clique with local interactions (colored spots). The insets illustrate that this diagonalization is achieved via a unitary that prepares a W-state over the atoms in the clique. To prepare a uniform $W$ state of size $s$, the rotation angles should satisfy $\cos(\theta_1/2) = 1/\sqrt{s}$ for the first rotation and $\sin(\theta_m/2) = - 1/\sqrt{s+1-m}$ for $m>1$. After applying the walk operator associated with one tessellation, the atomic connectivity is dynamically reconfigured, and the walk proceeds with the next tessellation in the sequence.}
    \label{fig:implementation-proposal}
\end{figure*}

Quantum walks, the quantum analog of classical random walks, are a natural tool to design quantum algorithms for graph-based problems. 
Besides providing a universal model of quantum computation \cite{childs2013universal}, quantum walks are used as computational primitives for tasks such as search, optimization, and quantum simulations \cite{childs2004spatial,magniez2007quantum,ambainis2007quantum,marsh2020combinatorial,wang2019simulating,di2016quantum}. Consequently, the ability to implement quantum walks over a large class of graphs is particularly important to be able to tackle more general network problems.

There are multiple experimental demonstrations of quantum walks on the line and on a square lattice using superconducting qubits \cite{yan2019strongly,gong2021quantum}, photonic chips \cite{tang2018experimental,chen2018observation,peruzzo2010quantum}, trapped ions \cite{zahringer2010realization,schmitz2009quantum,tamura2020quantum,matjeschk2012experimental,xue2009quantum} and cold atoms \cite{young2022tweezer,karski2009quantum,dadras2019experimental,preiss2015strongly,clark2021quantum}. However, a general method for realizing quantum walks on arbitrary graphs remains an open challenge due to the connectivity demanded by these graphs. Despite these advancements, these quantum walks are limited to specific graph structures, restricting their application to general graph problems. To address this, we propose a general implementation of a staggered quantum walk using Rydberg atom arrays. Our approach is particularly suited for spatial networks, formally defined as graph with nodes embedded in a metric space, where connectivity is strictly local and constrained by a maximum spatial distance. We also introduce a classical preprocessing algorithm to find a tessellation cover of the graph, which enables the realization of staggered quantum walks.

This article is organized as follows. In Section \ref{sec:proposal}, we review the definition of a staggered quantum walk and present our implementation proposal. In Section \ref{sec:tessellation}, we introduce our algorithm for constructing a tessellation cover, a crucial step for implementing the staggered quantum walk, and benchmark its performance on random geometric graphs. In Section \ref{sec:approaches}, we compare our approach to alternative quantum walk implementations, namely a coined quantum walk and a continuous-time quantum walk. In Section \ref{sec:search}, we demonstrate that our implementation performance is consistent with the optimal quadratic speedup in spatial search problems. Finally, we summarize our findings and discuss future directions in Section \ref{sec:conclusions}.

\section{Staggered walk proposal}
\label{sec:proposal}
The staggered model \cite{portugal2016staggered} is a discrete-time quantum walk that does not require a coin. Instead, its evolution is defined through the notion of graph tessellations. This family of quantum walks encompasses Szegedy's walks, a quantization of bipartite Markov chains \cite{szegedy2004quantum}. These models have attracted considerable interest due to their feasibility for physical implementation \cite{portugal2017staggered,khatibi2017staggered} and have also inspired new problems in graph theory \cite{abreu2017bounds,abreu2018graph,abreu2021computational}.

As illustrated in Fig.~\ref{fig:tessellation-cover}, a tessellation is a partition of the vertices into cliques, while a tessellation cover is a set of tessellations that cover all the edges of the graph. Each clique $\alpha_k$ in tessellation $\alpha$ is associated with a uniform superposition state of its vertices, $\ket{\alpha_k}$, which is used to define the reflection operator
\begin{equation}
    W_\alpha = 1 - 2 \sum_k \ket{\alpha_k} \bra{\alpha_k}.
\end{equation}
The overall walk operator is given by the product of the reflection operators for all tessellations, $U=W_\alpha W_\beta\dots$. 

To implement a staggered quantum walk in a graph with $N$ vertices, we use $N$ atoms with an excitation encoding, that is, the walker is a hyperfine excitation $\ket{i}=\ket{0_1 \dots 1_i \dots 0_N}$. 
While this requires more physical qubits than a compact $\mathcal{O}(\log N)$ binary encoding, it allows the graph topology to map directly onto the native geometry and interactions of the hardware. Then we diagonalize each walk operator
\begin{equation}
    W_\alpha = \bigotimes_{\alpha_k\in \alpha} U_{\alpha_k}\ \mathrm{C}^{s-1}\mathrm{Z}\ U_{\alpha_k}^\dagger,
\end{equation}
where $s=|\alpha_k|$ is the size of the $\alpha_k$ clique, C$^{s-1}$Z is a multi-controlled Z-gate acting on all qubits of the clique and $U_{\alpha_k}$ maps the state $\ket{1_1 1_2\dots 1_s}$ to the state $\ket{\alpha_k}$, a superposition of all single-excitation states within the clique. Since $\ket{\alpha_k}$ is a W-state,
\begin{equation}
\ket{\alpha_k}=\frac{\ket{1_1 0_2\dots 0_s} + \ket{0_1 1_2 \dots 0_s}+ \dots +\ket{0_1 0_2 \dots 1_s}}{\sqrt{s}},    
\end{equation}
$U_{\alpha_k}$ has a well-known circuit decomposition with $O(s)$ two-qubit gates, shown in Fig.~\ref{fig:implementation-proposal}. Therefore, the walk operator of each tessellation can be implemented with $O(N)$ gates. Additionally, our implementation makes use of the native multiqubit gates, $\mathrm{C}^{s-1}\mathrm{Z}$, of the Rydberg platform~\cite{morgado2021quantum}, which can be realized directly on neutral-atom arrays through Rydberg blockade and related multiqubit pulse protocols. In this setting, a global laser drive can condition the phase on the fully excited computational state without decomposing the operation into a deep sequence of two-qubit gates.
Alternative approaches based on Rydberg dressing~\cite{young2021asymmetric}, geometric phases~\cite{li2021multiple,pelegri2022high,rej2026neutral}, or antiblockade~\cite{su2017fast,su2018onestep} have also been developed, offering different trade-offs between robustness, interaction range, and spatial constraints. Similarly, the conditional $R_y$ rotations within each clique can be readily implemented without further rearrangements, but with the requirement for atom-resolved Rydberg excitation within all cliques.

\begin{algorithm}[alg:tessellations]{Tessellation cover}

\textbf{Main Routine}

\vspace{0.5em}
\textbf{Input:} Vertex $u$
\begin{enumerate}[leftmargin=*, label=\arabic*., itemsep=1pt, parsep=0pt]
    \item \textbf{For each} neighbor $v$ of $u$:
    \item \hspace{1em} \textbf{For each} color $c$ in \texttt{used\_colors}:
    \item \hspace{2em} \textbf{If} \texttt{is\_colorable}($u$, $v$, $c$):
    \item \hspace{3em} \texttt{color\_edges}($u$, $v$, $c$)
    \item \hspace{3em} \textbf{Continue} to next neighbor
    \item \hspace{1em} \textbf{If} no color $c$ worked:
    \item \hspace{2em} $c_{\text{new}} \gets$ \texttt{next\_color}()
    \item \hspace{2em} \texttt{color\_edges}($u$, $v$, $c_{\text{new}}$)
\end{enumerate}

\vspace{1em}
\hrule
\vspace{1em}

\textbf{Function} \texttt{is\_colorable}($u, v, c$)
\begin{enumerate}[leftmargin=*, label=\arabic*., itemsep=1pt, parsep=0pt]
    \item \textbf{For each} $x$ neighbor of $v$ \textbf{with} color $(v,x) = c$:
    \item \hspace{1em} \textbf{If} $(u,x)$ not an edge: \textbf{Return} False
    \item \textbf{For each} $y$ neighbor of $u$ \textbf{with} color $(u,y) = c$:
    \item \hspace{1em} \textbf{If} $(v,y)$ not an edge: \textbf{Return} False
    \item \textbf{Return} True
\end{enumerate}

\vspace{1em}
\hrule
\vspace{1em}

\textbf{Function} \texttt{color\_edges}($u, v, c$)
\begin{enumerate}[leftmargin=*, label=\arabic*., itemsep=1pt, parsep=0pt]
    \item Set color $(u,v) \gets c$
    \item \textbf{For each} $x$ neighbor of $v$ \textbf{with} color $(v,x) = c$:
    \item \hspace{1em} Set color $(u,x) \gets c$
    \item \textbf{For each} $y$ neighbor of $u$ \textbf{with} color $(u,y) = c$:
    \item \hspace{1em} Set color $(v,y) \gets c$
    \item \textbf{For each} $x$ neighbor of $v$ \textbf{with} color $(v,x) = c$:
    \item \hspace{1em} \textbf{For each} $y$ neighbor of $u$ \textbf{with} color $(u,y) = c$:
    \item \hspace{2em} Set color $(x,y) \gets c$
\end{enumerate}

\end{algorithm}

\section{Tessellation algorithm}
\label{sec:tessellation}
Since the staggered quantum walk relies on the tessellation structure, efficiently computing the tessellation cover is essential for scaling the model to large graphs. We developed an algorithm to find a tessellation cover in $O(m d^2)$ time, where $m$ is the number of edges in the graph and $d$ is the maximum degree. The algorithm starts with an empty graph and adds one vertex at a time, updating the tessellation cover according to Algorithm \ref{alg:tessellations}. This routine ensures the graph remains properly tessellated as new vertices and edges are added. For simplicity, we refer to tessellations as ``colors'', and ``coloring an edge'' means assigning it to a specific tessellation.

The computational cost of the algorithm is dominated by the \texttt{color\_edges} function, which is called exactly once per edge. Within this function, a nested loop iterates over the neighbors of a given vertex, yielding a complexity of $\mathcal{O}(m d^2)$. In contrast, the aggregate cost of the \texttt{is\_colorable} checks introduces a complexity of $\mathcal{O}(N d^2)$, which remains subdominant to the primary $\mathcal{O}(m d^2)$ bottleneck. To guarantee these bounds, we assume that checking edge existence and retrieving the neighbors of a vertex $v$ with a specific color $c$ are both $\mathcal{O}(1)$ operations rather than linear scans. In practice, this is achieved by using hash-set adjacency lists augmented with per-vertex, per-color neighbor buckets; specifically, by maintaining a nested hash map where \texttt{color\_class[v][c]} returns the set of neighbors $x$ satisfying $\text{color}(v,x) = c$.

Since the implementation depth is proportional to the size of the tessellation cover (i.e., the number of tessellations $T$), understanding how $T$ scales is crucial. 
Furthermore, characterizing the distribution of clique sizes within these tessellations is essential, as the clique size dictates the maximum number of qubits involved in the multi-qubit entangling gates required by the staggered quantum walk.
To investigate this, we study random geometric graphs $\operatorname{RGG}(N,r)$, which consist of $N$ random points uniformly distributed in the square $[0,1]^2$. Two points $x$ and $y$ are connected if their Euclidean distance satisfies $||x-y||<r$. There is a critical radius $r_c=\sqrt{\frac{\log N}{\pi N}}$ above which the graph is almost surely connected \cite{penrose1997longest,appel2002connectivity}. We set the natural scaling of $r$ to be the critical radius for connectivity $r_c$ and apply our tessellation algorithm. As a result, we find that the number of tessellations scales as $T = O \left(r/r_c \log N\right)$, as shown in Fig.~\ref{fig:tessellation-scaling}. In Appendix~\ref{sec:benchmark-heuristic}, we benchmark our heuristic algorithm against the exact minimum tessellation cover obtained through exhaustive enumeration of all possible tessellations. For the system sizes accessible to exact optimization ($N\leq 8$), we find that the heuristic produces tessellation covers that are close to optimal.

\begin{figure}
    \centering
    \includegraphics[width=0.99\linewidth]{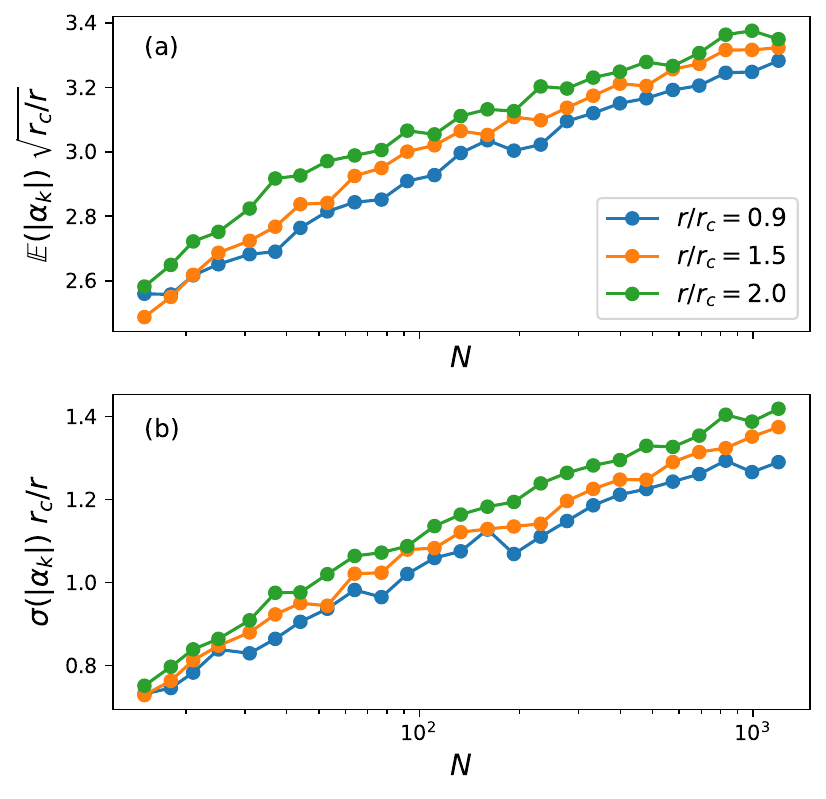}
    \caption{\justifying
    Scaling behavior of the clique-size distribution in random geometric graphs with $N$ vertices, showing (a) the scaled average and (b) the scaled standard deviation. The approximate collapse of the curves to a straight line implies that the average clique size scales as $\mathcal{O}(\sqrt{r/r_c} \log N)$, whereas the standard deviation scales as $\mathcal{O}\left(r/r_c \log N\right)$. The moments were computed from $6000/N$ independent graph realizations.}
    \label{fig:clique-size-scaling}
\end{figure}

By analyzing the distribution of non-trivial clique sizes (i.e., discarding 1-cliques), we find that the average scales as $\mathcal{O}(\sqrt{r/r_c} \log N)$, whereas the standard deviation scales as $\mathcal{O}\left(r/r_c \log N\right)$ (see Fig.~\ref{fig:clique-size-scaling}). Crucially, for fixed $r/r_c$, the standard deviation grows with the same dependence on $N$ as the mean. Consequently, the relative fluctuations remain finite in the large $N$ limit, and the distribution does not
concentrate around its average. Instead, it exhibits persistent fluctuations. This implies that the relevant resource for an experimental implementation is not only the average clique size but also the tail of the distribution.

Current neutral-atom platforms have demonstrated multi-qubit entangling operations involving several atoms, with three-qubit gates already experimentally realized and larger native multi-qubit phase operations being an active area of investigation~ \cite{su2017fast,su2018onestep,young2021asymmetric,li2021multiple,pelegri2022high,rej2026neutral}. Consequently, clique sizes in the range of a few atoms (e.g., $\lesssim 4-6$) are compatible with near-term neutral-atom implementations, whereas substantially larger cliques would likely require additional gate decomposition or optimization of the tessellation procedure. For the graph sizes and interaction radii considered here, the average clique sizes remain in this experimentally relevant regime. However, the persistence of fluctuations in the clique-size distribution indicates that rare large cliques can appear as the graph size increases. Consequently, for larger instances it may become necessary to introduce additional tessellation constraints or decompose large cliques into sequences of smaller multi-qubit operations. This trade-off between tessellation depth and experimentally feasible gate size represents an important consideration for scaling the implementation.

Finally, since $m = N d/2$ and $d = O((r/r_c)^2 \log N)$, the complexity of the tessellation algorithm is $O\left((r/r_c)^6 N \log^3{N}\right)$. This favorable scaling is a consequence of the geometric structure of spatial networks. In particular, the spatial embedding imposes a bound on the typical number of neighbors of each vertex, keeping the degree $d$ small compared to the total number of vertices. For graph classes without geometric constraints, the degree can grow significantly with the system size. For example, dense graphs may have $d=O(N)$, in which case the same algorithmic bound becomes $O(mN^2)$ and loses its practical advantage. Conversely, graph families with bounded degree, such as sparse networks, are expected to retain favorable scaling similar to the spatial networks studied here.

\begin{figure}
    \centering
    \includegraphics{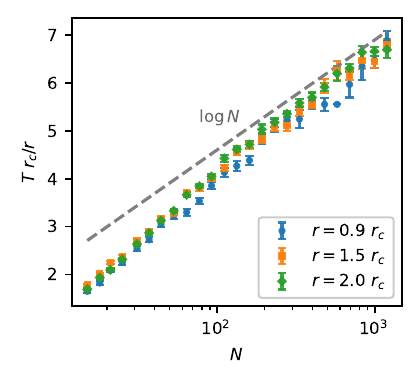}
    \caption{\justifying Scaling of the number of tessellations in random geometric graphs with $N$ vertices. The plot shows that our algorithm produces a number of tessellations approximately given by $T \approx (r/r_c) \log N$. The data points represent the mean ($\mu$), while the error bars indicate the standard error of the mean ($\sigma / \sqrt{M}$), where $\mu$ and $\sigma$ are computed over $M$ random realizations, with the number of realizations set to $M=6000/N$.
}
    \label{fig:tessellation-scaling}
\end{figure}

\begin{figure*}[t]
    \centering
    \raisebox{0.15\height}{\includegraphics[width=0.25\linewidth]{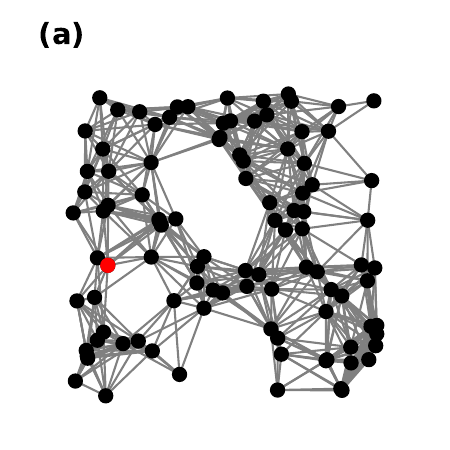}}
    \hfill
    \includegraphics[width=0.33\linewidth]{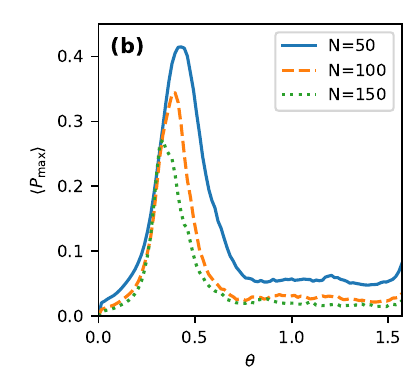}
    \hfill
    \includegraphics[width=0.33\linewidth]{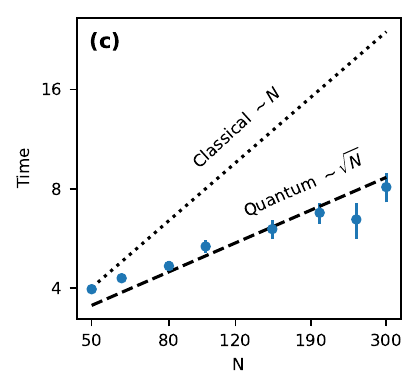}

    \caption{\justifying Spatial search with staggered quantum walk on random geometric graphs with $r/r_c=2$ (an example is shown on panel (a)). First, a scanning of $\theta$ is made, characterizing what are the values of $\theta$ that maximize the probability amplification (panel (b)). Then, we perform search with this value of $\theta$ and identify the time (i.e., number of oracle calls or, equivalently, staggered-walk iterations) taken to amplify the amplitude of the marked vertex. The results (panel (c), in log-log scale) show that the staggered quantum walk performance is consistent with a quadratic speedup. We use $10^4/N$ random geometric graphs per each graph size $N$.}
    \label{fig:spatial-search}
\end{figure*}

\section{Overview of other approaches} 
\label{sec:approaches}
There are other ideas to implement quantum walks over more general graphs. In Ref.~\cite{chakrabarti2012design}, the authors propose a generic implementation for discrete-time quantum walks based on the coined quantum walk model. In this approach, the degree of the graph is regularized so that a global coin can be used and the shift operators become permutations of the walker states, controlled by the coin register. However, the degree regularization introduces artificial self-loops and this can lead to a biased evolution and localization of the walker.

A different approach is a continuous-time quantum walk (CTQW), where the evolution operator is given by $U(t)=e^{-i H t}$ and $H$ is taken to be proportional to the adjacency matrix of the graph. The na\"ive approach to implement a CTQW is to exactly decompose $U(t)$ into single and two-qubit gates. This can be achieved with a quantum circuit of $n=\lceil\log_2 N\rceil$ qubits and $O(N^2)$ gates \cite{mottonen12006decompositions}. The decomposition is based on a QR factorization of the matrix $U(t)$, whose complexity is $O(N^3)$ \cite{vartiainen2004efficient}.

Alternatively, we can approximate the evolution via Trotterization \cite{nielsen2001quantum}. Since the Hamiltonian can be written as a sum of two-body interactions
\begin{equation}
    H=\sum_{i<j} h_{ij} = \sum_{i<j} \gamma \left(\ket{i}\bra{j} + \ket{j} \bra{i}\right),
\end{equation}
where $\gamma>0$ is a constant, the time evolution can be broken into a large number $K=t/\Delta t$ of steps, such that we only need two-body interactions at a time:
\begin{equation}
    e^{-i H \Delta t} = \prod_{i<j} e^{-i h_{ij} \Delta t} + O(\Delta t^2).
\end{equation}

To implement this first-order Trotter formula, we can use an array of $N$ Rydberg atoms. The walker is encoded as an excitation $\ket{i}=\ket{0_1 \dots 1_i \dots 0_N}$. To perform $e^{-i h_{ij} \Delta t}$, we send the states $\ket{0}$ and $\ket{1}$ to dipole-coupled Rydberg states (say with angular momentum $s$ and $p$) and let the atoms evolve through 
\begin{equation}
    h^\text{Ryd}_{ij} = \frac{C_3}{R_{ij}^3} \left(\ket{s_i p_j}\bra{p_i s_j} + \text{h.c.}\right),
\end{equation}
where $R_{ij}$ is the distance between atom $i$ and $j$ \cite{morgado2021quantum}. The number of steps $K$ depends on the desired accuracy of the approximation, $\epsilon$, and it is proportional to $t^2/\epsilon$ \cite{lloyd1996universal}. Due to the locality of spatial networks, we can assume that atoms corresponding to adjacent vertices are close enough to neglect any transport time. This way, each interaction time is approximately just $\Delta t=t/K \sim \epsilon/t$. Since there are $m t^2/\epsilon$ interactions, where $m$ is the number of edges, the overall running time is proportional to $m t^2/\epsilon$. A major drawback of this approach is its inefficiency with respect to the accuracy $\epsilon$, the dependence on $1/\epsilon$ instead of $1/\log(\epsilon)$ indicates that to get an extra bit of precision we need to double the running time.

\section{Spatial search}
\label{sec:search} 
To demonstrate the potential of our proposal, we benchmark its performance on a spatial search algorithm. The goal of spatial search is to find a marked vertex in a graph using as few queries as possible. Similarly to the Grover's search algorithm, spatial search is built on two key ingredients: an oracle, which marks the target vertex by flipping its phase, and a diffusion operator, which coherently spreads amplitudes across the graph. Repeated application of these two operators leads to an amplification of the probability of the marked vertex. The number of oracle calls required to find the target is referred to as the query complexity. In classical algorithms, searching an unstructured graph typically requires checking every vertex, leading to a search time that scales with the number of vertices $N$. In contrast, quantum algorithms can achieve a quadratic speedup, reducing the number of queries to $O(\sqrt{N})$.

The oracle is implemented as a phase flip on the marked vertex, labeled $\ket{0}$:
\begin{equation}
    R = 1- 2 \ket{0}\bra{0}.
\end{equation}
For the diffusion operator, we adopt the strategy of Ref. \cite{portugal2017quantum} and use a generalized staggered quantum walk, which is defined as
\begin{equation}
    W_\theta = e^{-i \theta W_1} e^{-i \theta W_2}\dots e^{-i \theta W_T},
    \label{eq:generalized-staggered-walk}
\end{equation}
where $W_\alpha =1-2 \sum_k \ket{\alpha_k}\bra{\alpha_k}$ are the staggered walk operators. The parameter $\theta$ represents the rotation angle of the walk and can be tuned to maximize the efficiency of the search. When $\theta=\pi/2$, we recover the original staggered walk (up to a global phase). Implementing this generalized staggered walk \cite{portugal2017staggered} requires replacing the C$^{s-1}$Z gate with a C$^{s-1}$Z$_\theta$ gate, which is also a native Rydberg gate \cite{yu2022multiqubit,he2022multiple}.

To assess its performance, we simulate the algorithm on random geometric graphs with $r/r_c=2$ and periodic boundary conditions, to mitigate finite-size effects. First, we determine the value of $\theta$ that produces the highest amplitude amplification of the marked vertex with a linear scan (see Appendix~\ref{sec:numerical-procedure}). Then, using this optimal value $\theta = \theta_\text{op}$, we measured the search time, defined as the number of oracle calls needed to reach the first maximum in probability. As shown in Fig.~\ref{fig:spatial-search}, the search time is consistent with $\mathcal{O}(\sqrt{N})$, while the number of tessellations steps is $\mathcal{O}(T \sqrt{N}) = \mathcal{O}(\sqrt{N} \log N )$. Possible reasons for the small deviations to the $\mathcal{O}(\sqrt{N})$ asymptotic behaviour include finite-$N$ corrections and the quality of the tessellation cover produced by our heuristic algorithm.

\section{Conclusions}
\label{sec:conclusions}
In this work, we proposed a novel implementation of quantum walks using Rydberg atoms. Our implementation encodes the walker as an excitation and diagonalizes the walk operators of a staggered quantum walk, using Rydberg multiqubit gates and exploiting the platform's capability to dynamically rearrange atoms. This implementation is particularly well-suited for spatial networks, which exhibit non-trivial connectivity and model many real-world systems. Additionally, we developed an efficient classical algorithm to find a tessellation cover and benchmarked it on random geometric graphs. Finally, we demonstrated that our proposal performance is consistent with a quadratic speedup in a spatial search algorithm.

While our protocol is platform-independent in principle, Rydberg atoms would offer unique tailored advantages. Their native planar connectivity and ability to dynamically rearrange atoms allow the system to reconfigure the interactions for each tessellation step of the walk. Furthermore, Rydberg interactions allow the direct implementation of multiqubit gates (e.g., through blockade mechanisms), avoiding the overhead of decomposition into two-qubit gates. As the size of graphs increase, some cliques may exceed the experimentally feasible size for a single multiqubit gate. In such cases, we could either decompose it into smaller multiqubit gates or adapt the tessellation algorithm to restrict the maximum clique size. 

Regarding the quantum search algorithm, although we showed the existence of an optimal angle that enables a speedup, determining this value directly from the graph remains an open question. Developing an efficient method to estimate this angle would significantly enhance the algorithm's applicability. Alternatively, $\theta$ can be treated as a variational parameter, optimized classically using a feedback loop, as proposed in the Quantum Walk Optimization Algorithm (QWOA) framework \cite{marsh2019quantum}.

Our work opens several avenues for future exploration. An experimental realization of this proposal would allow for the validation of our theoretical findings and investigate the effect of noise and decoherence. While our approach is a step towards more versatile quantum walk implementations, developing efficient methods for a wider class of non-trivial graphs, in particular non-local networks, remains an important challenge that could unlock further application of quantum technologies.

\section{Acknowledgments}
The authors thank Mario Szegedy and Shannon Whitlock for valuable discussions. The authors gratefully acknowledge the support from FCT -- Fundação para a Ciência e a Tecnologia (Portugal), namely through project UIDB/ 04540/2020, as well as from project EuRyQa – European infrastructure for Rydberg Quantum Computing (GA 101070144) of the Horizon Europe Programme of the European Commission. The authors acknowledge the computational resources of CNCA funded by FCT.

\section{Author Contributions}

Y.O. conceptualized and initialized the idea. G.A. proposed the Rydberg implementation, developed the tessellation algorithm and performed all numerical simulations. All authors discussed the results, analyzed the data, and contributed to writing of the manuscript. Additionally, a Large Language Model (ChatGPT) was used to provide language support, and the resulting text was subsequently reviewed, edited, and approved by all authors, who take full responsibility for the integrity and content of the work.

\bibliographystyle{quantum}

\begin{thebibliography}{10}

\bibitem{posfai2016network}
Albert-László Barabási and Márton Pósfai.
\newblock ``Network science''.
\newblock \href{https://dx.doi.org/10.1098/rsta.2012.0375}{Volume~3}.
\newblock Cambridge University Press. ~(2016).

\bibitem{fortunato2010community}
Santo Fortunato.
\newblock ``Community detection in graphs''.
\newblock \href{https://dx.doi.org/10.1016/j.physrep.2009.11.002}{Physics Reports {\bf 486}, 75--174}~(2010).

\bibitem{danila2006optimal}
Bogdan Danila, Yong Yu, John~A. Marsh, and Kevin~E. Bassler.
\newblock ``Optimal transport on complex networks''.
\newblock \href{https://dx.doi.org/10.1103/PhysRevE.74.046106}{Phys. Rev. E {\bf 74}, 046106}~(2006).

\bibitem{yan2006efficient}
Gang Yan, Tao Zhou, Bo~Hu, Zhong-Qian Fu, and Bing-Hong Wang.
\newblock ``Efficient routing on complex networks''.
\newblock \href{https://dx.doi.org/10.1103/PhysRevE.73.046108}{Phys. Rev. E {\bf 73}, 046108}~(2006).

\bibitem{moreno2004dynamics}
Yamir Moreno, Maziar Nekovee, and Amalio~F. Pacheco.
\newblock ``Dynamics of rumor spreading in complex networks''.
\newblock \href{https://dx.doi.org/10.1103/PhysRevE.69.066130}{Phys. Rev. E {\bf 69}, 066130}~(2004).

\bibitem{morgado2021quantum}
M.~Morgado and S.~Whitlock.
\newblock ``{Quantum simulation and computing with Rydberg-interacting qubits}''.
\newblock \href{https://dx.doi.org/10.1116/5.0036562}{AVS Quantum Science {\bf 3}, 023501}~(2021).

\bibitem{mcinroy2024benchmarking}
K~McInroy, N~Pearson, and JD~Pritchard.
\newblock ``Benchmarking the algorithmic performance of near-term neutral atom processors''~(2024).
\newblock  url:~\url{https://doi.org/10.48550/arXiv.2402.02127}.

\bibitem{maskara2025programmable}
Nishad Maskara, Stefan Ostermann, James Shee, Marcin Kalinowski, Abigail McClain~Gomez, Rodrigo Araiza~Bravo, Derek~S Wang, Anna~I Krylov, Norman~Y Yao, Martin Head-Gordon, et~al.
\newblock ``Programmable simulations of molecules and materials with reconfigurable quantum processors''.
\newblock \href{https://dx.doi.org/10.1038/s41567-024-02738-z}{Nature Physics {\bf 21}, 289--297}~(2025).

\bibitem{ebadi2022quantum}
S.~Ebadi, A.~Keesling, M.~Cain, T.~T. Wang, H.~Levine, D.~Bluvstein, G.~Semeghini, A.~Omran, J.-G. Liu, R.~Samajdar, X.-Z. Luo, B.~Nash, X.~Gao, B.~Barak, E.~Farhi, S.~Sachdev, N.~Gemelke, L.~Zhou, S.~Choi, H.~Pichler, S.-T. Wang, M.~Greiner, V.~Vuletić, and M.~D. Lukin.
\newblock ``{Quantum optimization of maximum independent set using Rydberg atom arrays}''.
\newblock \href{https://dx.doi.org/10.1126/science.abo6587}{Science {\bf 376}, 1209--1215}~(2022).

\bibitem{kim2022rydberg}
Minhyuk Kim, Kangheun Kim, Jaeyong Hwang, Eun-Gook Moon, and Jaewook Ahn.
\newblock ``Rydberg quantum wires for maximum independent set problems''.
\newblock \href{https://dx.doi.org/10.1038/s41567-022-01629-5}{Nature Physics {\bf 18}, 755--759}~(2022).

\bibitem{jeong2023quantum}
Seokho Jeong, Minhyuk Kim, Minki Hhan, JuYoung Park, and Jaewook Ahn.
\newblock ``{Quantum programming of the satisfiability problem with Rydberg atom graphs}''.
\newblock \href{https://dx.doi.org/10.1103/PhysRevResearch.5.043037}{Phys. Rev. Res. {\bf 5}, 043037}~(2023).

\bibitem{young2022tweezer}
Aaron~W. Young, William~J. Eckner, Nathan Schine, Andrew~M. Childs, and Adam~M. Kaufman.
\newblock ``{Tweezer-programmable 2D quantum walks in a Hubbard-regime lattice}''.
\newblock \href{https://dx.doi.org/10.1126/science.abo0608}{Science {\bf 377}, 885--889}~(2022).

\bibitem{harrigan2021quantum}
Matthew~P Harrigan, Kevin~J Sung, Matthew Neeley, Kevin~J Satzinger, Frank Arute, Kunal Arya, Juan Atalaya, Joseph~C Bardin, Rami Barends, Sergio Boixo, et~al.
\newblock ``Quantum approximate optimization of non-planar graph problems on a planar superconducting processor''.
\newblock \href{https://dx.doi.org/10.1038/s41567-020-01105-y}{Nature Physics {\bf 17}, 332--336}~(2021).

\bibitem{zhu2022multi}
Yingyue Zhu, Zewen Zhang, Bhuvanesh Sundar, Alaina~M Green, C~Huerta Alderete, Nhung~H Nguyen, Kaden~RA Hazzard, and Norbert~M Linke.
\newblock ``Multi-round {QAOA} and advanced mixers on a trapped-ion quantum computer''.
\newblock \href{https://dx.doi.org/10.1088/2058-9565/ac91ef}{Quantum science and technology {\bf 8}, 015007}~(2022).

\bibitem{deng2023solving}
Yu-Hao Deng, Si-Qiu Gong, Yi-Chao Gu, Zhi-Jiong Zhang, Hua-Liang Liu, Hao Su, Hao-Yang Tang, Jia-Min Xu, Meng-Hao Jia, Ming-Cheng Chen, Han-Sen Zhong, Hui Wang, Jiarong Yan, Yi~Hu, Jia Huang, Wei-Jun Zhang, Hao Li, Xiao Jiang, Lixing You, Zhen Wang, Li~Li, Nai-Le Liu, Chao-Yang Lu, and Jian-Wei Pan.
\newblock ``{Solving Graph Problems Using Gaussian Boson Sampling}''.
\newblock \href{https://dx.doi.org/10.1103/PhysRevLett.130.190601}{Phys. Rev. Lett. {\bf 130}, 190601}~(2023).

\bibitem{childs2013universal}
Andrew~M. Childs, David Gosset, and Zak Webb.
\newblock ``{Universal Computation by Multiparticle Quantum Walk}''.
\newblock \href{https://dx.doi.org/10.1126/science.1229957}{Science {\bf 339}, 791--794}~(2013).

\bibitem{childs2004spatial}
Andrew~M. Childs and Jeffrey Goldstone.
\newblock ``Spatial search by quantum walk''.
\newblock \href{https://dx.doi.org/10.1103/PhysRevA.70.022314}{Phys. Rev. A {\bf 70}, 022314}~(2004).

\bibitem{magniez2007quantum}
Fr{\'e}d{\'e}ric Magniez, Miklos Santha, and Mario Szegedy.
\newblock ``{Quantum Algorithms for the Triangle Problem}''.
\newblock \href{https://dx.doi.org/10.1137/050643684}{SIAM Journal on Computing {\bf 37}, 413--424}~(2007).

\bibitem{ambainis2007quantum}
Andris Ambainis.
\newblock ``{Quantum Walk Algorithm for Element Distinctness}''.
\newblock \href{https://dx.doi.org/10.1137/S0097539705447311}{SIAM Journal on Computing {\bf 37}, 210--239}~(2007).

\bibitem{marsh2020combinatorial}
S.~Marsh and J.~B. Wang.
\newblock ``Combinatorial optimization via highly efficient quantum walks''.
\newblock \href{https://dx.doi.org/10.1103/PhysRevResearch.2.023302}{Phys. Rev. Res. {\bf 2}, 023302}~(2020).

\bibitem{wang2019simulating}
Kunkun Wang, Xingze Qiu, Lei Xiao, Xiang Zhan, Zhihao Bian, Wei Yi, and Peng Xue.
\newblock ``{Simulating Dynamic Quantum Phase Transitions in Photonic Quantum Walks}''.
\newblock \href{https://dx.doi.org/10.1103/PhysRevLett.122.020501}{Phys. Rev. Lett. {\bf 122}, 020501}~(2019).

\bibitem{di2016quantum}
Giuseppe Di~Molfetta and Armando P{\'e}rez.
\newblock ``Quantum walks as simulators of neutrino oscillations in a vacuum and matter''.
\newblock \href{https://dx.doi.org/10.1088/1367-2630/18/10/103038}{New J. of Phys. {\bf 18}, 103038}~(2016).

\bibitem{yan2019strongly}
Zhiguang Yan, Yu-Ran Zhang, Ming Gong, Yulin Wu, Yarui Zheng, Shaowei Li, Can Wang, Futian Liang, Jin Lin, Yu~Xu, Cheng Guo, Lihua Sun, Cheng-Zhi Peng, Keyu Xia, Hui Deng, Hao Rong, J.~Q. You, Franco Nori, Heng Fan, Xiaobo Zhu, and Jian-Wei Pan.
\newblock ``Strongly correlated quantum walks with a 12-qubit superconducting processor''.
\newblock \href{https://dx.doi.org/10.1126/science.aaw1611}{Science {\bf 364}, 753--756}~(2019).

\bibitem{gong2021quantum}
Ming Gong, Shiyu Wang, Chen Zha, Ming-Cheng Chen, He-Liang Huang, Yulin Wu, Qingling Zhu, Youwei Zhao, Shaowei Li, Shaojun Guo, Haoran Qian, Yangsen Ye, Fusheng Chen, Chong Ying, Jiale Yu, Daojin Fan, Dachao Wu, Hong Su, Hui Deng, Hao Rong, Kaili Zhang, Sirui Cao, Jin Lin, Yu~Xu, Lihua Sun, Cheng Guo, Na~Li, Futian Liang, V.~M. Bastidas, Kae Nemoto, W.~J. Munro, Yong-Heng Huo, Chao-Yang Lu, Cheng-Zhi Peng, Xiaobo Zhu, and Jian-Wei Pan.
\newblock ``Quantum walks on a programmable two-dimensional 62-qubit superconducting processor''.
\newblock \href{https://dx.doi.org/10.1126/science.abg7812}{Science {\bf 372}, 948--952}~(2021).

\bibitem{tang2018experimental}
Hao Tang, Xiao-Feng Lin, Zhen Feng, Jing-Yuan Chen, Jun Gao, Ke~Sun, Chao-Yue Wang, Peng-Cheng Lai, Xiao-Yun Xu, Yao Wang, Lu-Feng Qiao, Ai-Lin Yang, and Xian-Min Jin.
\newblock ``Experimental two-dimensional quantum walk on a photonic chip''.
\newblock \href{https://dx.doi.org/10.1126/sciadv.aat3174}{Science Advances {\bf 4}, eaat3174}~(2018).

\bibitem{chen2018observation}
Chao Chen, Xing Ding, Jian Qin, Yu~He, Yi-Han Luo, Ming-Cheng Chen, Chang Liu, Xi-Lin Wang, Wei-Jun Zhang, Hao Li, Li-Xing You, Zhen Wang, Da-Wei Wang, Barry~C. Sanders, Chao-Yang Lu, and Jian-Wei Pan.
\newblock ``{Observation of Topologically Protected Edge States in a Photonic Two-Dimensional Quantum Walk}''.
\newblock \href{https://dx.doi.org/10.1103/PhysRevLett.121.100502}{Phys. Rev. Lett. {\bf 121}, 100502}~(2018).

\bibitem{peruzzo2010quantum}
Alberto Peruzzo, Mirko Lobino, Jonathan C.~F. Matthews, Nobuyuki Matsuda, Alberto Politi, Konstantinos Poulios, Xiao-Qi Zhou, Yoav Lahini, Nur Ismail, Kerstin Wörhoff, Yaron Bromberg, Yaron Silberberg, Mark~G. Thompson, and Jeremy~L. OBrien.
\newblock ``{Quantum Walks of Correlated Photons}''.
\newblock \href{https://dx.doi.org/10.1126/science.1193515}{Science {\bf 329}, 1500--1503}~(2010).

\bibitem{zahringer2010realization}
F.~Z{\"a}hringer, G.~Kirchmair, R.~Gerritsma, E.~Solano, R.~Blatt, and C.~F. Roos.
\newblock ``{Realization of a Quantum Walk with One and Two Trapped Ions}''.
\newblock \href{https://dx.doi.org/10.1103/PhysRevLett.104.100503}{Phys. Rev. Lett. {\bf 104}, 100503}~(2010).

\bibitem{schmitz2009quantum}
H.~Schmitz, R.~Matjeschk, Ch. Schneider, J.~Glueckert, M.~Enderlein, T.~Huber, and T.~Schaetz.
\newblock ``{Quantum Walk of a Trapped Ion in Phase Space}''.
\newblock \href{https://dx.doi.org/10.1103/PhysRevLett.103.090504}{Phys. Rev. Lett. {\bf 103}, 090504}~(2009).

\bibitem{tamura2020quantum}
Masaya Tamura, Takashi Mukaiyama, and Kenji Toyoda.
\newblock ``{Quantum Walks of a Phonon in Trapped Ions}''.
\newblock \href{https://dx.doi.org/10.1103/PhysRevLett.124.200501}{Phys. Rev. Lett. {\bf 124}, 200501}~(2020).

\bibitem{matjeschk2012experimental}
Robert Matjeschk, Ch~Schneider, Martin Enderlein, Thomas Huber, Hector Schmitz, Jan Glueckert, and Tobias Schaetz.
\newblock ``Experimental simulation and limitations of quantum walks with trapped ions''.
\newblock \href{https://dx.doi.org/10.1088/1367-2630/14/3/035012}{New J. Phys. 14 {\bf 14}, 035012}~(2012).

\bibitem{xue2009quantum}
Peng Xue, Barry~C. Sanders, and Dietrich Leibfried.
\newblock ``{Quantum Walk on a Line for a Trapped Ion}''.
\newblock \href{https://dx.doi.org/10.1103/PhysRevLett.103.183602}{Phys. Rev. Lett. {\bf 103}, 183602}~(2009).

\bibitem{karski2009quantum}
Michal Karski, Leonid Förster, Jai-Min Choi, Andreas Steffen, Wolfgang Alt, Dieter Meschede, and Artur Widera.
\newblock ``{Quantum Walk in Position Space with Single Optically Trapped Atoms}''.
\newblock \href{https://dx.doi.org/10.1126/science.1174436}{Science {\bf 325}, 174--177}~(2009).

\bibitem{dadras2019experimental}
Siamak Dadras, Alexander Gresch, Caspar Groiseau, Sandro Wimberger, and Gil~S. Summy.
\newblock ``Experimental realization of a momentum-space quantum walk''.
\newblock \href{https://dx.doi.org/10.1103/PhysRevA.99.043617}{Phys. Rev. A {\bf 99}, 043617}~(2019).

\bibitem{preiss2015strongly}
Philipp~M. Preiss, Ruichao Ma, M.~Eric Tai, Alexander Lukin, Matthew Rispoli, Philip Zupancic, Yoav Lahini, Rajibul Islam, and Markus Greiner.
\newblock ``Strongly correlated quantum walks in optical lattices''.
\newblock \href{https://dx.doi.org/10.1126/science.1260364}{Science {\bf 347}, 1229--1233}~(2015).

\bibitem{clark2021quantum}
J.~H. Clark, C.~Groiseau, Z.~N. Shaw, S.~Dadras, C.~Binegar, S.~Wimberger, G.~S. Summy, and Y.~Liu.
\newblock ``Quantum to classical walk transitions tuned by spontaneous emissions''.
\newblock \href{https://dx.doi.org/10.1103/PhysRevResearch.3.043062}{Phys. Rev. Res. {\bf 3}, 043062}~(2021).

\bibitem{portugal2016staggered}
Renato Portugal.
\newblock ``Staggered quantum walks on graphs''.
\newblock \href{https://dx.doi.org/10.1103/PhysRevA.93.062335}{Phys. Rev. A {\bf 93}, 062335}~(2016).

\bibitem{szegedy2004quantum}
M.~Szegedy.
\newblock ``Quantum speed-up of markov chain based algorithms''.
\newblock In 45th Annual IEEE Symposium on Foundations of Computer Science.
\newblock \href{https://dx.doi.org/10.1109/FOCS.2004.53}{Pages 32--41}.
\newblock ~(2004).

\bibitem{portugal2017staggered}
R.~Portugal, M.~C. de~Oliveira, and J.~K. Moqadam.
\newblock ``{Staggered quantum walks with Hamiltonians}''.
\newblock \href{https://dx.doi.org/10.1103/PhysRevA.95.012328}{Phys. Rev. A {\bf 95}, 012328}~(2017).

\bibitem{khatibi2017staggered}
J.~Khatibi~Moqadam, M.~C. de~Oliveira, and R.~Portugal.
\newblock ``Staggered quantum walks with superconducting microwave resonators''.
\newblock \href{https://dx.doi.org/10.1103/PhysRevB.95.144506}{Phys. Rev. B {\bf 95}, 144506}~(2017).

\bibitem{abreu2017bounds}
A~Abreu, L~Cunha, T~Fernandes, C~de~Figueiredo, L~Kowada, F~Marquezino, D~Posner, and R~Portugal.
\newblock ``Bounds and complexity for the tessellation problem''.
\newblock \href{https://dx.doi.org/10.21711/231766362017/rmc453}{Mat. Contemp {\bf 45}, 22--30}~(2017).

\bibitem{abreu2018graph}
Alexandre Abreu, Luis Cunha, Tharso Fernandes, Celina de~Figueiredo, Luis Kowada, Franklin Marquezino, Daniel Posner, and Renato Portugal.
\newblock ``{The Graph Tessellation Cover Number: Extremal Bounds, Efficient Algorithms and Hardness}''.
\newblock In LATIN 2018: Theoretical Informatics.
\newblock \href{https://dx.doi.org/10.1007/978-3-319-77404-6\_1}{Pages 1--13}.
\newblock Cham~(2018). Springer International Publishing.

\bibitem{abreu2021computational}
Alexandre Abreu, Luís Cunha, Celina {de Figueiredo}, Luis Kowada, Franklin Marquezino, Renato Portugal, and Daniel Posner.
\newblock ``A computational complexity comparative study of graph tessellation problems''.
\newblock \href{https://dx.doi.org/10.1016/j.tcs.2020.11.045}{Theoretical Computer Science {\bf 858}, 81--89}~(2021).

\bibitem{young2021asymmetric}
Jeremy~T. Young, Przemyslaw Bienias, Ron Belyansky, Adam~M. Kaufman, and Alexey~V. Gorshkov.
\newblock ``{Asymmetric Blockade and Multiqubit Gates via Dipole-Dipole Interactions}''.
\newblock \href{https://dx.doi.org/10.1103/PhysRevLett.127.120501}{Phys. Rev. Lett. {\bf 127}, 120501}~(2021).

\bibitem{li2021multiple}
Meng Li, F.-Q. Guo, Z.~Jin, L.-L. Yan, E.-J. Liang, and S.-L. Su.
\newblock ``{Multiple-qubit controlled unitary quantum gate for Rydberg atoms using shortcut to adiabaticity and optimized geometric quantum operations}''.
\newblock \href{https://dx.doi.org/10.1103/PhysRevA.103.062607}{Phys. Rev. A {\bf 103}, 062607}~(2021).

\bibitem{pelegri2022high}
Gerard Pelegr{\'\i}, Andrew~J Daley, and Jonathan~D Pritchard.
\newblock ``{High-fidelity multiqubit Rydberg gates via two-photon adiabatic rapid passage}''.
\newblock \href{https://dx.doi.org/10.1088/2058-9565/ac823a}{Quantum Science \& Technology {\bf 7}, 045020}~(2022).

\bibitem{rej2026neutral}
Sinchan~Snigdha Rej and Bimalendu Deb.
\newblock ``Neutral-atom quantum computation using multi-qubit geometric gates via adiabatic passage''.
\newblock \href{https://dx.doi.org/10.1088/1367-2630/ae390c}{New J. Phys. {\bf 28}, 014513}~(2026).

\bibitem{su2017fast}
Shi-Lei Su, Ya~Gao, Erjun Liang, and Shou Zhang.
\newblock ``{Fast Rydberg antiblockade regime and its applications in quantum logic gates}''.
\newblock \href{https://dx.doi.org/10.1103/PhysRevA.95.022319}{Phys. Rev. A {\bf 95}, 022319}~(2017).

\bibitem{su2018onestep}
S.~L. Su, H.~Z. Shen, Erjun Liang, and Shou Zhang.
\newblock ``{One-step construction of the multiple-qubit Rydberg controlled-phase gate}''.
\newblock \href{https://dx.doi.org/10.1103/PhysRevA.98.032306}{Phys. Rev. A {\bf 98}, 032306}~(2018).

\bibitem{penrose1997longest}
Mathew~D Penrose.
\newblock ``The longest edge of the random minimal spanning tree''.
\newblock \href{https://dx.doi.org/10.1214/aoap/1034625335}{Ann. Appl. Probab. {\bf 7}, 340--361}~(1997).

\bibitem{appel2002connectivity}
Martin~J.B. Appel and Ralph~P. Russo.
\newblock ``The connectivity of a graph on uniform points on $[0,1]^d$''.
\newblock \href{https://dx.doi.org/10.1016/S0167-7152(02)00233-X}{Statistics \& Probability Letters {\bf 60}, 351--357}~(2002).

\bibitem{chakrabarti2012design}
Amlan Chakrabarti, ChiaChun Lin, and Niraj~K. Jha.
\newblock ``{Design of Quantum Circuits for Random Walk Algorithms}''.
\newblock In 2012 IEEE Computer Society Annual Symposium on VLSI.
\newblock \href{https://dx.doi.org/10.1109/ISVLSI.2012.45}{Pages 135--140}.
\newblock ~(2012).

\bibitem{mottonen12006decompositions}
Mikko M{\"o}tt{\"o}nen and Juha~J Vartiainen.
\newblock ``Decompositions of general quantum gates''.
\newblock \href{https://dx.doi.org/10.48550/arXiv.quant-ph/0504100}{Chapter~7}.
\newblock Nova Publishers. ~(2006).

\bibitem{vartiainen2004efficient}
Juha~J. Vartiainen, Mikko M{\"o}tt{\"o}nen, and Martti~M. Salomaa.
\newblock ``{Efficient Decomposition of Quantum Gates}''.
\newblock \href{https://dx.doi.org/10.1103/PhysRevLett.92.177902}{Phys. Rev. Lett. {\bf 92}, 177902}~(2004).

\bibitem{nielsen2001quantum}
Michael~A. Nielsen and Isaac~L. Chuang.
\newblock ``{Quantum Computation and Quantum Information}''.
\newblock \href{https://dx.doi.org/10.1017/CBO9780511976667}{Volume~2}.
\newblock Cambridge University Press. ~(2001).

\bibitem{lloyd1996universal}
Seth Lloyd.
\newblock ``{Universal Quantum Simulators}''.
\newblock \href{https://dx.doi.org/10.1126/science.273.5278.1073}{Science {\bf 273}, 1073--1078}~(1996).

\bibitem{portugal2017quantum}
R.~Portugal and T.~D. Fernandes.
\newblock ``{Quantum search on the two-dimensional lattice using the staggered model with Hamiltonians}''.
\newblock \href{https://dx.doi.org/10.1103/PhysRevA.95.042341}{Phys. Rev. A {\bf 95}, 042341}~(2017).

\bibitem{yu2022multiqubit}
Dongmin Yu, Han Wang, Jin-ming Liu, Shi-Lei Su, Jing Qian, and Weiping Zhang.
\newblock ``{Multiqubit Toffoli Gates and Optimal Geometry with Rydberg Atoms}''.
\newblock \href{https://dx.doi.org/10.1103/PhysRevApplied.18.034072}{Phys. Rev. Applied {\bf 18}, 034072}~(2022).

\bibitem{he2022multiple}
Yucheng He, Jing-Xin Liu, F.-Q. Guo, L.-L. Yan, Ronghui Luo, Erjun Liang, S.-L. Su, and M.~Feng.
\newblock ``{Multiple-qubit Rydberg quantum logic gate via dressed-state scheme}''.
\newblock \href{https://dx.doi.org/10.1016/j.optcom.2021.127500}{Optics Communications {\bf 505}, 127500}~(2022).

\bibitem{marsh2019quantum}
S.~Marsh and J.B. Wang.
\newblock ``{A quantum walk-assisted approximate algorithm for bounded NP optimisation problems}''.
\newblock \href{https://dx.doi.org/10.1007/s11128-019-2171-3}{Quantum Inf Process {\bf 18}, 61}~(2019).

\end{thebibliography}

\appendix

\section{Benchmarking heuristic against minimal tessellation cover}
\label{sec:benchmark-heuristic}

To assess the quality of the greedy tessellation-cover algorithm, we
compare its output against the exact minimum tessellation cover obtained
through exhaustive search on small random geometric graphs. To this end,
we reformulate the tessellation-cover problem as a finite set-cover
problem. The exact algorithm consists of three main stages: (i) exhaustive
generation of all possible tessellations, (ii) conversion of each
tessellation into the set of edges that it covers, and (iii) an exact
search for the smallest collection of tessellations whose covered-edge
sets reproduce the full edge set of the graph.

The first step is to generate all possible partitions of the vertex set
into cliques. This is performed using a recursive backtracking procedure.
At each step of the recursion, an unassigned vertex $v$ is selected
according to a fixed ordering. The algorithm then enumerates every
possible clique containing $v$ whose vertices belong to the remaining
unassigned set. Each such clique defines one possible block of the
partition. The algorithm removes the selected clique from the remaining
vertices and repeats the procedure until all vertices have been assigned.

This generates the complete set of tessellations
\begin{equation}
\mathcal{T}=\{T_1,T_2,\ldots,T_M\},
\end{equation}
where each tessellation satisfies
\begin{equation}
\bigcup_{C\in T_i} C = V,
\end{equation}
and each clique $C\in T_i$ induces a complete subgraph of $G$.
The fixed ordering of vertices removes redundant permutations of the same
partition and ensures that each tessellation is generated exactly once.

For each tessellation $T_i$, we define the set of edges covered by that
tessellation as
\begin{equation}
E_i =
\left\{
(u,v)\in E :
\exists C\in T_i \text{ such that } u,v\in C
\right\}.
\end{equation}
The tessellation-cover problem is therefore equivalent to finding the
smallest subset of tessellations
\begin{equation}
\{T_{i_1},T_{i_2},\ldots,T_{i_k}\}
\end{equation}
such that
\begin{equation}
E_{i_1}\cup E_{i_2}\cup\cdots\cup E_{i_k}=E .
\end{equation}

This is precisely a minimum set-cover problem, with universe $E$ and
available sets $\{E_1,\ldots,E_M\}$. The first value of $k$ for which such
a collection exists defines the optimal tessellation cover number.

The minimum tessellation cover is obtained by searching over increasing
values of $k$. For each value, the algorithm performs a recursive
backtracking search over possible collections of $k$ tessellations. At some recursion depth, the algorithm keeps track of the already covered edges $E_{\mathrm{cov}}$ and the remaining uncovered edges $E_{\mathrm{uncov}}=E\setminus E_{\mathrm{cov}}$. A branch terminates successfully whenever $E_{\mathrm{uncov}}=\emptyset$, indicating that a valid tessellation cover has been found. Since smaller values of $k$ are tested first, the first successful solution is guaranteed to be optimal.

Several optimizations are employed to reduce the computational cost.
First, tessellations are sorted according to the number of edges they cover,
with tessellations covering more edges considered earlier. This increases
the probability of finding a solution early in the search.

Second, permutation symmetry among selected tessellations is removed by
requiring their indices to satisfy $i_1\leq i_2\leq\cdots\leq i_k$. Thus, different orderings of the same set of tessellations are not explored
separately.

Finally, the search is pruned whenever the remaining tessellations cannot
possibly cover the remaining uncovered edges. If the next admissible
tessellation index is $i$, the backtrack branch is discarded whenever
\begin{equation}
E_{\mathrm{uncov}}
\not\subseteq
\bigcup_{j\geq i}E_j .
\end{equation}

\begin{figure}[ht]
    \centering
    \includegraphics[width=\linewidth]{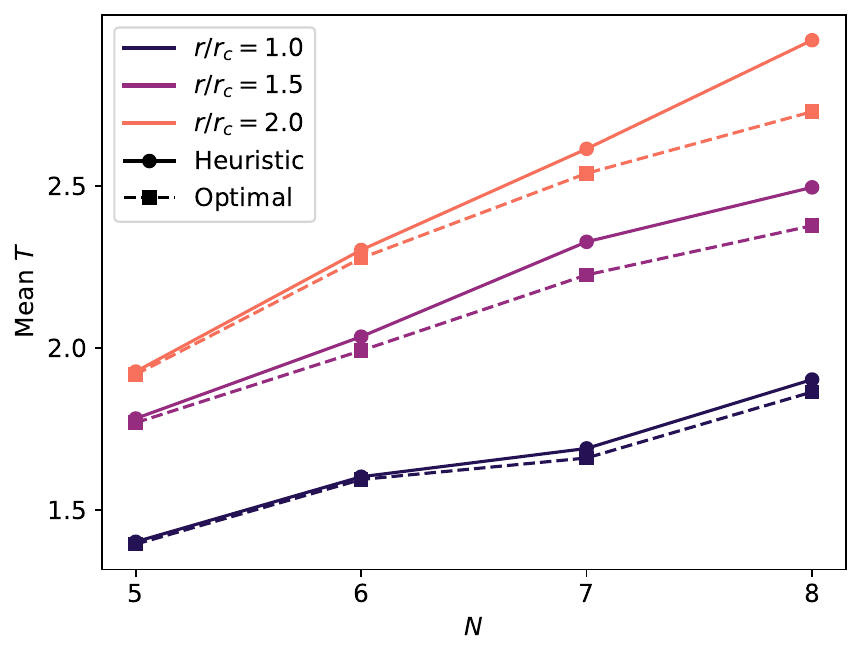}
    \caption{
        Mean number of tessellations produced by the greedy algorithm
        (circles) and by exhaustive optimization (squares) as a function of
        graph size $N$, for different values of $r/r_c$.
    }
    \label{fig:exhaustive_comparison}
\end{figure}

We generate random geometric graphs with $N$ vertices distributed
uniformly in the unit square and with connectivity radius $r$. For each
graph instance, we compute both the exact tessellation-cover number and the number of tessellations obtained by the
greedy algorithm. Fig.~\ref{fig:exhaustive_comparison} compares the mean
values of these two quantities.

For the range of system sizes accessible to exhaustive optimization, the
greedy algorithm produces covers close to the optimal solution, with only a
small increase in the average number of tessellations relative to the
minimum cover. The computational limitations of exhaustive search restrict
the comparison to relatively small systems, preventing conclusions about
the asymptotic scaling of the deviation from optimality.

\section{Numerical implementation of spatial search}
\label{sec:numerical-procedure}

In this appendix we describe the numerical procedure used to obtain the results in
Fig.~\ref{fig:spatial-search}. For each system size $N$, we generate random
geometric graphs $\operatorname{RGG}(N,r)$ by placing $N$ vertices uniformly at
random in the unit square $[0,1]^2$. Two vertices are connected whenever their
Euclidean distance is smaller than the interaction radius $r$. Periodic boundary
conditions are imposed when computing distances in order to reduce finite-size and boundary
effects. The interaction radius is set to $r=2r_c$. For each graph instance, the tessellation cover is constructed using the greedy algorithm described in Sec.~\ref{sec:tessellation}. 

The resulting cover defines the generalized staggered quantum walk operator $W_\theta$, in Eq.~\eqref{eq:generalized-staggered-walk}, where $\theta$ is the reflection angle associated with each tessellation. The search operator is then constructed as $U = W_\theta R$, where the oracle $R$ marks vertex $0$ by applying a phase flip.

The initial state is the uniform superposition over all vertices,
\begin{equation}
    |\psi_0\rangle =
\frac{1}{\sqrt{N}}\sum_{j=0}^{N-1}|j\rangle .
\end{equation}
For each value of $\theta$, the evolution is simulated by iterating the search
operator 50 times and recording the probability of measuring the marked vertex,
\begin{equation}
P_{\mathrm{m}}(t)=|\langle 0|\psi(t)\rangle|^2 .    
\end{equation}
The angle $\theta$ is optimized by performing a uniform scan over the interval
$[0,\pi/2]$ with 100 values and selecting the value that maximizes the first
peak probability. For a given graph instance, the search time is defined as the
first time at which the marked-vertex probability exceeds $0.9$ of its maximum value over the simulation time,
\begin{equation}
    t_{\mathrm{op}}=\min\{t:P_{\mathrm{m}}(t)\geq 0.9\,\max_t P_{\mathrm{m}}(t)\}.
\end{equation}
This definition provides a robust estimate of the time required to reach the first amplification peak, taking into account the discrete-step structure of the walk.

For each value of $N$, the above procedure is repeated over an ensemble of
random graph instances. The number of graph realizations is chosen as
$10\,000/N$, keeping the total computational effort approximately constant
across different system sizes. The reported search times correspond to the mean
over the ensemble,
\begin{equation}
\overline{t}=\frac{1}{M}\sum_{i=1}^{M}t_i ,    
\end{equation}
with error bars given by the standard error of the mean,
\begin{equation}
\sigma_{\overline{t}}=\frac{\sigma_t}{\sqrt{M}},
\end{equation}
where $M$ is the number of graph realizations.

The observed scaling is therefore obtained from an ensemble average over random
geometric graphs, with the same tessellation construction and optimization
procedure applied independently to each realization.

\end{document}